\documentclass[final,5p,times,twocolumn,authoryear]{elsarticle}
\usepackage{amssymb}
\usepackage{graphicx}
\usepackage{array}
\usepackage{xcolor}
\usepackage{ulem}

\journal{Journal of High Energy Astrophysics}
\begin{document}
\begin{frontmatter}

\title{Detection of time delay between UV and X-ray variability in Mrk 1044 using \textit{AstroSat} observations}

\author[first]{M. Reshma}
\author[second]{C. S. Stalin}
\author[third]{Amit Kumar Mandal}
\author[second]{Abhijit Kayal}
\author[first]{S. B. Gudennavar \thanks{E-mail: shivappa.b.gudennavar@christuniversity.in}}
\author[second]{Prajwel Joseph}

\affiliation[first]{organization={Department of Physics and Electronics, CHRIST University}, 
            city={Bangalore 560029}, state={Karnataka}, country={India}}

\affiliation[second]{organization={Indian Institute of Astrophysics}, 
            addressline={Block II, Koramangala}, city={Bangalore 560034}, state={Karnataka}, country={India}}

\affiliation[third]{organization={Center for Theoretical Physics of the Polish Academy of Sciences}, 
            addressline={Al. Lotniko´w 32/46}, city={02–668 Warsaw}, country={Poland}}
     
\begin{abstract}
Active galactic nuclei are known to exhibit flux variations across the entire electromagnetic spectrum. Among these, correlations between UV/optical and X-ray flux variations serve as a key diagnostics for understanding the physical connection between the accretion disk and the corona. In this work, we present the results of analysis of ultraviolet (UV) and X-ray flux variations in the narrow line Seyfert 1 galaxy Mrk 1044. Simultaneous observations in the far-UV band (FUV: 1300$-$1800 \AA) and the X-ray band (0.5$-$7 keV) obtained during 31 August $-$ 8 September 2018 with the Ultraviolet Imaging Telescope and the Soft X-ray Telescope onboard \textit{AstroSat} were used for this study. Significant flux variability was detected in both FUV and X-ray bands. The  fractional root mean square variability amplitude ($F_{\rm var}$) was found to be 0.036 $\pm$ 0.001 in the FUV band and 0.384 $\pm$ 0.004 in the X-ray band. To explore potential time lag between the two bands, cross-correlation analysis was performed using both the interpolated cross-correlation function (ICCF) and just another vehicle for estimating lags in nuclei (JAVELIN) methods. Results from both approaches are consistent within 2$\sigma$ uncertainty, indicating that X-ray variations lead the FUV variations, with measured lags of 2.25$\pm$0.05 days (ICCF) and $2.35_{-0.01}^{+0.02}$ days (JAVELIN). This is the first detection of a time delay between UV and X-ray variations in Mrk 1044. The observed UV lag supports the disk reprocessing scenario, wherein X-ray emission from the corona irradiates the accretion disk, driving the observed UV variability.
\end{abstract}

\end{frontmatter}

\section{Introduction}
\label{introduction}
Active galactic nuclei (AGN) are among the most luminous sources in the Universe, with bolometric luminosities in the range $10^{40}$ $-$ $10^{47}$ erg s$^{-1}$. They are thought to be powered by accretion of matter onto supermassive black holes (SMBHs: $10^{6}$ $-$ $10^{10}$ M$_\odot$) located at the centres of galaxies \citep{lynden1969galactic,rees1984black,fabian1999active}. According to one of the standard accretion disk models \citep{shakura1973black}, the infalling material forms an optically thick, geometrically thin disk around the SMBH, which emits thermal radiation primarily in the ultraviolet (UV)/optical bands. However, the observed spectral energy distribution of AGN is more complex than the multi-temperature blackbody spectrum as predicted by the standard accretion disk model.
\\[6pt]
The majority of AGN are radio-quiet \citep{kellermann1994radio}, producing little to no radio emission. Their variable X-ray emission is generally attributed to the inverse Compton scattering of UV/optical photons from the accretion disk by hot ($\sim$$10^{9}$ K) electrons in a compact corona located near the disk \citep{haardt1993x, haardt1994model}. Although the physical nature and geometry of the corona remain subjects of debate, recent X-ray polarimetric observations with \textit{Imaging X-ray Polarimetry Explorer (IXPE)} have provided new insights. These observations do not favour a simple spherical lamp-post geometry, instead suggest an extended slab or wedge-shaped corona in NGC 4151 \citep{gianolli2023uncovering} and a cone-shaped geometry in IC 4329A \citep{pal2023x}.
\\[6pt]
Regardless of its geometry, inverse Compton scattering produces a power law X-ray spectrum with a characteristic high-energy cutoff. A fraction of this emission directly reaches the observer, while the remaining portion irradiates the accretion disk. In the latter case, part of the incident radiation is absorbed, heating the disk and thereby enhancing its UV/optical emission, while the rest gives rise to the reflection component observed in the X-ray spectrum. In this framework, any variability in the X-ray emission is expected to drive correlated variations in the reprocessed UV/optical emission from the disk \citep{clavel1992correlated}, with the latter lagging behind the X-rays by a timescale consistent with the disk reprocessing model \citep{krolik1991ultraviolet, cackett2007testing}.
\\[6pt]
One of the defining characteristics of AGN is their flux variability across the entire electromagnetic spectrum \citep{ulrich1997variability}. In many AGN, flux variations observed at different wavelengths are found to be correlated, offering key insights into the physical coupling between different emission regions. Of particular interest is the correlation between UV and X-ray flux variability, as this provides crucial clues to the complex interplay between the accretion disk and the corona. If the observed UV variability is primarily driven by X-ray reprocessing, the X-ray flux variations are expected to lead the UV variations \citep{krolik1991ultraviolet, cackett2007testing}. Conversely, if the X-ray variations lag behind the UV variations, it indicates that the X-rays originate from inverse Compton scattering of UV/optical seed photons from the disk, supporting the thermal Comptonization model \citep{adegoke2019uv}. Therefore, measuring time lags between UV and X-ray flux variations serves as a powerful diagnostic for probing the physical mechanisms and geometry of the innermost regions of AGN. However, we caution that the two processes, namely the X-ray emission from the corona and reprocessing/heating of the accretion disk are not independent, but strongly coupled. Because of this coupling, the observed time lags may not be a clear indication of a single process. Instead the observed time lag must be a combination of disk reprocessing and Comptonization processes. Thus depending on which process dominates over the other during the observing period and the energy bands used for the observations, one may get reduced lag, no lag or lag with complex energy dependence.
\\[6pt]
Results available in the literature on the correlation between UV and X-ray flux variations in AGN are, however, diverse and often contradictory. In several sources such as Ark 120 \citep{lobban2020x}, NGC 4593 \citep{kumari2023contrasting} and NGC 4051 \citep{kumari2024detection}, the X-ray variations are found to lead the UV variations, consistent with the X-ray reprocessing scenario, where X-rays incident on the accretion disk are reprocessed into UV/optical emission. In contrast, a few AGN including NGC 7469 \citep{kumari2023contrasting}, Mrk 493 \citep{adegoke2019uv}, MCG-6-30-15 \citep{arevalo2005x}, and NGC 6814 \citep{kumari2025exploring} exhibit X-ray variations that lag behind the UV variations, supporting the thermal Comptonization scenario, in which X-rays originate from inverse Compton scattering of UV/optical seed photons from the disk. Additionally, in Mrk 79, \citet{breedt2009long} reported that the UV and X-ray variations are correlated on timescales of days to a few tens of days, but with zero lag.  Thus, the current understanding of the UV and X-ray emission connection in AGN remains inconclusive, emphasizing the need for further studies involving simultaneous UV and X-ray monitoring of a larger number of sources. To this end, we are conducting a systematic investigation of UV and X-ray flux variability in a sample of AGN observed with \textit{AstroSat} \citep{agrawal2017astrosat}, which is uniquely suited for such studies due to its capability for simultaneous observations in both UV and X-ray bands. In this paper, we present our results for one such source, Mrk 1044.
\\[6pt]
Mrk 1044 is a narrow line Seyfert 1 galaxy hosting a SMBH of mass 2.8 $\times$ $10^{6}$ M$_\odot$ \citep{krongold2021detection} with mass accretion rate ranging from 1.2 \citep{husemann2022close} to 16 times the Eddington limit \citep{du2015supermassive}. The source has been extensively studied in the X-ray regime. Using a short 8 ks \textit{XMM-Newton} observation, \citet{dewangan2007investigation} found no evidence of a time lag between variations in the 0.2 $-$ 0.3 keV and 5 $-$ 10 keV bands, suggesting that reprocessing is unlikely to be responsible for the soft X-ray excess. Instead, they proposed that the excess could arise from inverse Compton scattering by thermal electrons. Subsequent spectral and timing analyses by \citet{mallick2018high} confirmed the presence of the soft X-ray excess, the Fe K$\alpha$ X-ray line, and a Compton hump in the 15 $-$ 30 keV range. A more recent study by \citet{barua2023search} also detected the soft X-ray excess and the Fe K$\alpha$ X-ray line but reported no strong correlation between the UV and X-ray light curves on either short or long timescales. The authors attributed this lack of correlation to the enhanced X-ray variability and strong general relativistic effects operating in the vicinity of the black hole.
\\[6pt]
In this work, we investigate the presence of time lag between UV and X-ray variations in Mrk 1044 using simultaneous FUV and X-ray observations obtained with \textit{AstroSat}. The paper is organized as follows: Section \ref{sec: observation and reductions} describes the observations and data reduction procedures; Section \ref{sec: analysis} presents the analysis of variability and the correlation between the UV and X-ray light curves; and Sections \ref{sec: discussion} and \ref{sec: summary} provide the discussion and conclusions, respectively. 

\section{Observation and data reduction}
\label{sec: observation and reductions}
\begin{figure}
\centering
\includegraphics[scale=0.31]{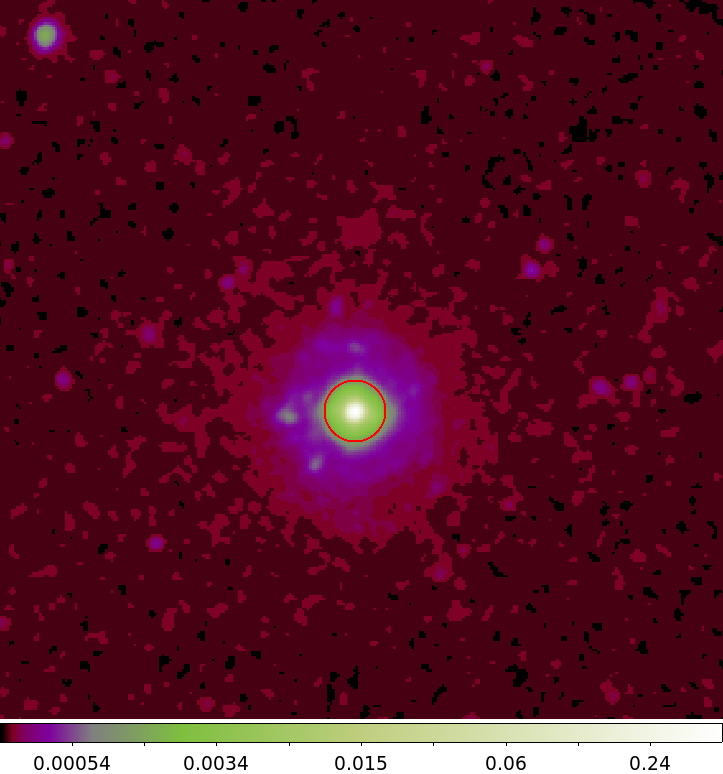}	
\caption{UVIT field image (2$\times$2 square arcmin) of Mrk 1044 in F154W filter. The circular aperture (of 5 arcsec radius) is marked in red. The colour scale is in units of counts/s.} 
\label{figure 1}
\end{figure}

\begin{figure}
\centering 
\includegraphics[scale=0.31]{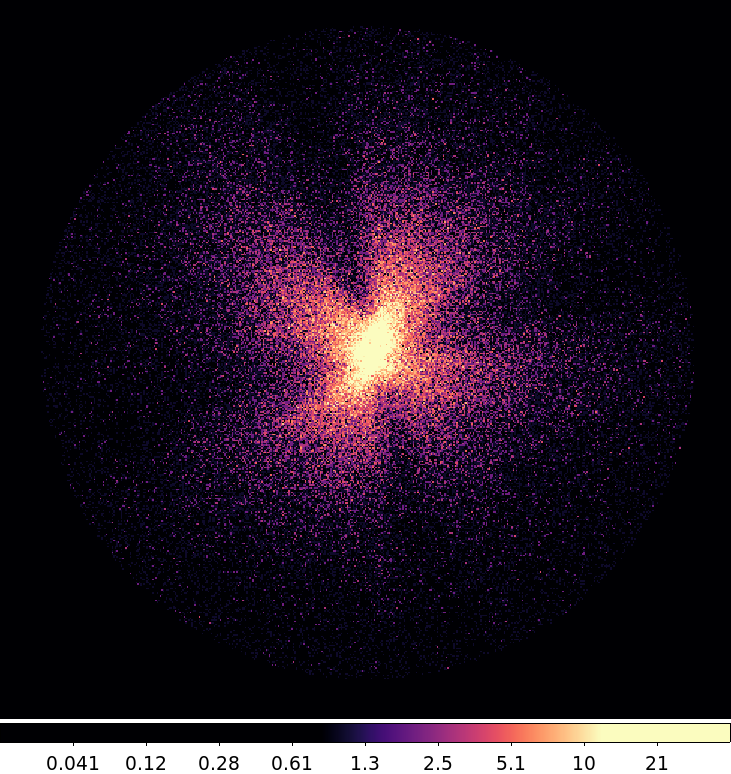}	
\caption{SXT field image of Mrk 1044 in the 0.5$-$7 keV energy range. The image is clipped to the 15 arcmin source region of interest with the detector corner regions excluded. The colour scale is in units of counts.} 
\label{figure 2}
\end{figure}

\begin{figure}
\centering 
\includegraphics[scale=0.51]{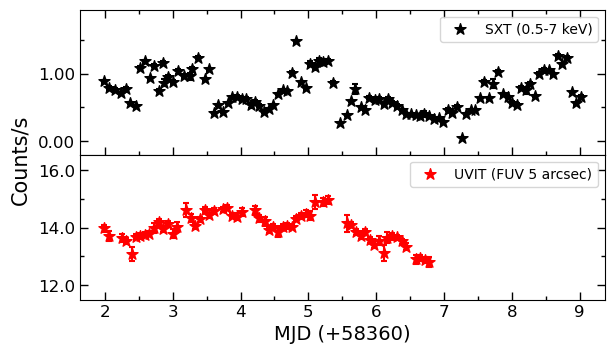}	
\caption{The light curves of Mrk 1044: X-ray in the 0.5$-$7 keV energy range (top panel) and FUV light curve using 5 arcsec aperture radius (bottom panel).} 
\label{figure 3}
\end{figure}

We used data obtained with the Ultraviolet Imaging Telescope (UVIT: \citealt{tandon2017orbit, tandon2020additional}) and the Soft X-ray Telescope (SXT: \citealt{singh2016orbit, singh2017soft}) onboard \textit{AstroSat} \citep{agrawal2017astrosat}, which was launched by the Indian Space Research Organization on 28 September 2015. The UVIT, with a circular field of view of $\sim$28 arcmin in diameter, is capable of imaging the sky simultaneously in the far-ultraviolet (FUV) and near-ultraviolet (NUV) channels, achieving a spatial resolution better than 1.5 arcsec. The SXT, though similar in design to \textit{Swift/XRT}, provides lower spatial resolution but is optimized for sensitive soft X-ray observations.
\\[6pt]
The \textit{AstroSat}/UVIT observed Mrk 1044 for a total exposure of 65 ks between 31 August 2018 to 5 September 2018 in the FUV band using the F154W filter (mean wavelength = 1541 \AA, bandwidth = 380 \AA) under Observation ID: A04\_143T01\_9000002344. For the present analysis, we used the Level-2 UVIT data processed with the UVIT pipeline version 7.0 \citep{joseph2025uvit}, obtained from the Indian Space Science Data Center (ISSDC)\footnote{https://www.issdc.gov.in/AS1\_Data\_Download.html}. We generated orbit-wise FUV light curve using the curve$\_$orbitwise() function in the {\scshape Curvit} Python package \citep{joseph2021curvit}, which operates on the calibrated events list. The source flux was extracted from a circular region of radius 12 subpixels ($\sim$5 arcsec), and aperture correction was applied following \citet{tandon2020additional} to obtain the total flux. The resulting light curve was further corrected for background and saturation effects. The FUV field image covering a 2$\times$2 square arcmin region, along with the aperture (circle of 5 arcsec radius) used for photometry, is shown in Fig. \ref{figure 1}. While the chosen 5 arcsec circular aperture encompasses some light from the host galaxy, the observed variation in the FUV light curve is driven solely by AGN variability and one does not expect variations from the host galaxy.  
\\[6pt]
The \textit{AstroSat}/SXT observed Mrk 1044 for a total exposure of 178 ks between 31 August 2018 and 8 September 2018 under the same Observation ID: A04\_143T01\_9000002344. Level-2 SXT data, processed with $\mathtt {sxtpipeline}$ (Version: 1.4b), downloaded from the ISSDC archive, were used along with the latest background spectrum file available from the SXT website\footnote{https://www.tifr.res.in/astrosat\_sxt/index.html}. The cleaned events files from multiple orbits were merged using the $\mathtt {sxtevtmergertool}$ in $\mathtt {Julia}$ (Version 1.10.0)\footnote{https://github.com/gulabd/SXTMerger.jl}.  The merged cleaned events file was then loaded into $\mathtt {XSELECT}$ (part of the $\mathtt {HEASOFT}$ package), from which light curve with a time resolution of 2.3775 s was extracted. A circular source region of radius 15 arcmin was chosen in the 0.5$-$7 keV energy range, as the detector response is uncertain below 0.5 keV and background-dominated beyond 7 keV. The light curve was rebinned orbit-wise using a bin size of 5820 s by determining the mean count rate for one orbit following the procedure outlined by \citet{kumari2023contrasting}. To obtain the background-subtracted source light curve, an average constant background rate for a region of 15 arcmin radius derived from the most recent background spectrum file, ‘SkyBkg\_sxt\_LE0p35\_R16p0\_v05\_Gd0to12.pha’ (release date: 2024-01-03) was used. The SXT field image in the 0.5$-$7 keV, with the 15 arcmin circular aperture used for photometry, is shown in Fig. \ref{figure 2}, and the simultaneous background subtracted FUV and X-ray light curves are displayed in Fig. \ref{figure 3}. The SXT spectrum for the same source region was extracted, which encloses 96\% of the encircled energy fraction. The spectrum was binned using the $\mathtt {ftgrouppha}$ task with grouptype = "opt", applying the optimal binning scheme of \citet{kaastra2016optimal}. A vignetting-corrected auxiliary response file (ARF) was generated using $\mathtt {sxtarfmodule}$, employing the recently updated ARF file ‘sxt\_pc\_excl00\_v04a\_20240917.arf’, which corrects for the sharp response jump near 1 keV.

\section{Flux variability, time lag and spectral analysis}
\label{sec: analysis}
\subsection{Flux variability}
To investigate flux variability in the FUV and X-ray bands, we calculated the fractional root mean square variability amplitude ($F_{\rm var}$), which measures the intrinsic source variance after correcting for measurement uncertainties. Following \citet{vaughan2003characterizing}, $F_{\rm var}$ is defined as: 

\begin{equation}
    F_{var} = \sqrt{\frac{S^2 - \overline{\sigma_{err}^2}}{\bar{x}^2}}
\end{equation}
where, $\bar{x}$ is the arithmetic mean of $x_{i}$, $S^2$ is sample variance, $\overline{\sigma_{err}^2}$ is the mean squared error. $S^2$ and $\overline{\sigma_{err}^2}$ are defined as: 

\begin{equation}
    S^2 = \frac{1}{n-1} \sum_{i=1}^n (x_{i}-\bar{x})^2
\end{equation}

\begin{equation}
    \overline{\sigma_{err}^2} = \frac{1}{n} \sum_{i=1}^n \sigma_{err,i}^2
\end{equation}
where, n is the number of data points in the light curve and $\sigma_{err,i}$ is the measured uncertainty on each data point. The error in $F_{var}$ is calculated as: 

\begin{equation}
    err (F_{var}) = \sqrt{\left(\sqrt{\frac{1}{2n}}\frac{\overline{\sigma_{err}^2}}{\bar{x}^2F_{var}}\right) ^2  + \left(\sqrt{\frac{\overline{\sigma_{err}^2}}{n}} \frac{1}{\bar{x}}\right)^2}
\end{equation} 
For the FUV light curve (duration = 64.8 ks), we obtained $F_{\rm var}^{\rm FUV}$ = 0.036 $\pm$ 0.001. For the X-ray light curve (duration = 177.8 ks), the corresponding value was $F_{\rm var}^{\rm SXT} = 0.384 \pm 0.004$. When considering only the SXT data overlapping with the UVIT data, we obtained $F_{\rm var}^{\rm SXT} = 0.370 \pm 0.004$. This indicates that the source is approximately ten times more variable in X-rays than in the FUV band. 

\subsection{Time lag}
\begin{figure}
\centering
\includegraphics[scale= 0.27]{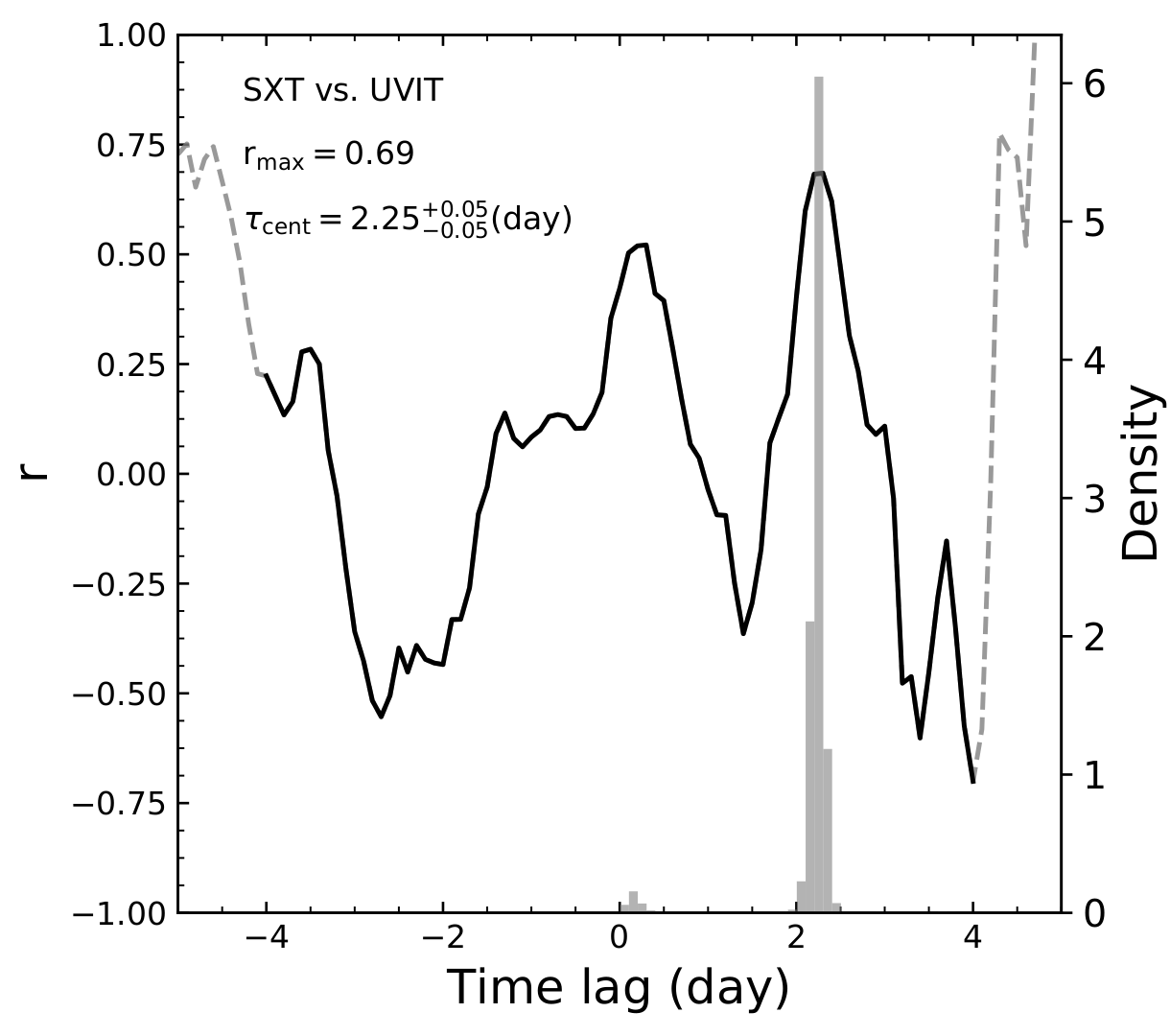}	
\caption{The cross-correlation function (solid line) between X-ray and FUV light curves. The histograms in grey are the distribution of the centroids of the cross-correlation function using the ICCF method. The dashed grey lines are the spurious peaks near the window boundaries attributed to aliasing in the light curves.} 
\label{figure 4}
\end{figure}

\begin{figure}
\includegraphics[scale= 0.20]{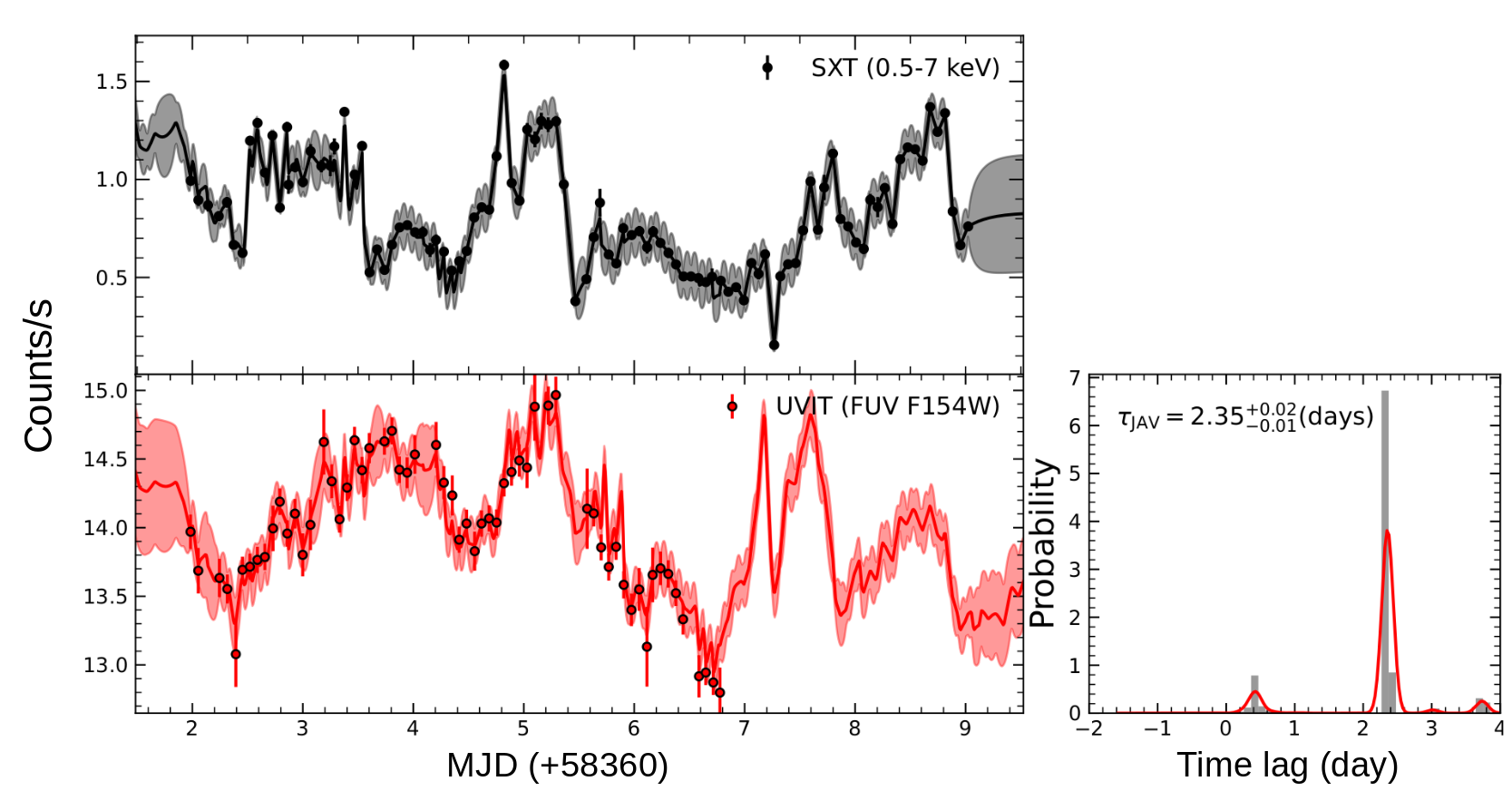}	
\caption{Left: X-ray light curve (top panel) and FUV light curve (bottom panel) of Mrk 1044. The solid lines show the JAVELIN-model fitted light curves, with uncertainties indicated by the shaded regions. Right: The histogram represents the lag probability distribution, with the smooth kernel density shown by the red line.}
\label{figure 5}
\end{figure}

The FUV and X-ray light curves cover durations of 4.8 days and 7 days, respectively. Accordingly, we carried out the time lag analysis over a range of $-$5 to +5 days using two standard techniques: the interpolated cross-correlation function (ICCF: \citealt{gaskell1986line,gaskell1987accuracy}) and just another vehicle for estimating lags in nuclei (JAVELIN: \citealt{zu2011alternative}). The ICCF results are presented in Fig. \ref{figure 4}. In this analysis, spurious peaks appearing near the window boundaries ($\pm$ 5 days, shown with dashed grey lines in Fig. \ref{figure 4}), likely caused by aliasing effects in the light curves \citep{grier2019sloan,homayouni2020sloan,mandal2024revisiting}, were excluded. The final lag and the associated uncertainties were determined via the cross-correlation centroid distribution. This was done by creating numerous randomized versions of the original light curves to build a distribution of centroid lags. The lag centroid was estimated from points exceeding 80\% of the ICCF peak value, using the flux randomization/random subset selection technique \citep{peterson1998uncertainties} with 4000 realizations. The median of the cross-correlation centroid distribution was adopted as the lag, and the 68\% confidence interval was taken as its uncertainty. However, we note that, time delays shorter than the value constrained by the present data may be revealed through observations with improved signal-to-noise ratio, enhanced temporal resolution and extended time coverage.
\\[6pt]
For the JAVELIN analysis, the X-ray continuum was modeled as a damped random walk (\citealt{kelly2009variations}) process. A top-hat transfer function was then convolved with the continuum to generate the UV response light curve, representing a shifted, scaled, and smoothed version of the driving signal. Model parameters and their posterior distributions were derived using Markov Chain Monte Carlo (MCMC) optimization. The resulting model fits and corresponding probability distributions are presented in Fig. \ref{figure 5}.
\\[6pt]
From the ICCF analysis, we obtained a median lag of 2.25$\pm$0.05 days between X-ray and FUV variations, indicating that the X-ray variations lead those in the FUV. Similarly, the JAVELIN analysis yielded a median lag of $2.35_{-0.01}^{+0.02}$ days, with the X-ray again leading the FUV variations. The lag estimates from both methods are found to be consistent within 2$\sigma$. There are spurious features in the lag distributions appearing at around 0.2 days in the CCF (Fig. \ref{figure 4}) and at approximately 0.4 days and 3.8 days in the JAVELIN lag probability distribution (Fig. \ref{figure 5}). These features can arise from correlated measurement noise introduced by the nearly identical sampling of the two light curves, as well as from the UV light curve retaining a blurred imprint of the X-ray variability, which can produce spurious peaks near zero lag. The peak near 3.8 days in the JAVELIN lag probability distribution may additionally be attributed to aliasing in the light curves. Importantly, these spurious contributions together account for only $\sim$4-5\% of the total lag probability across the full search window. Hence, their impact is negligible and does not affect the derived lag.

\subsection{Spectral analysis}
\begin{figure}
\centering
\includegraphics[scale=0.30]{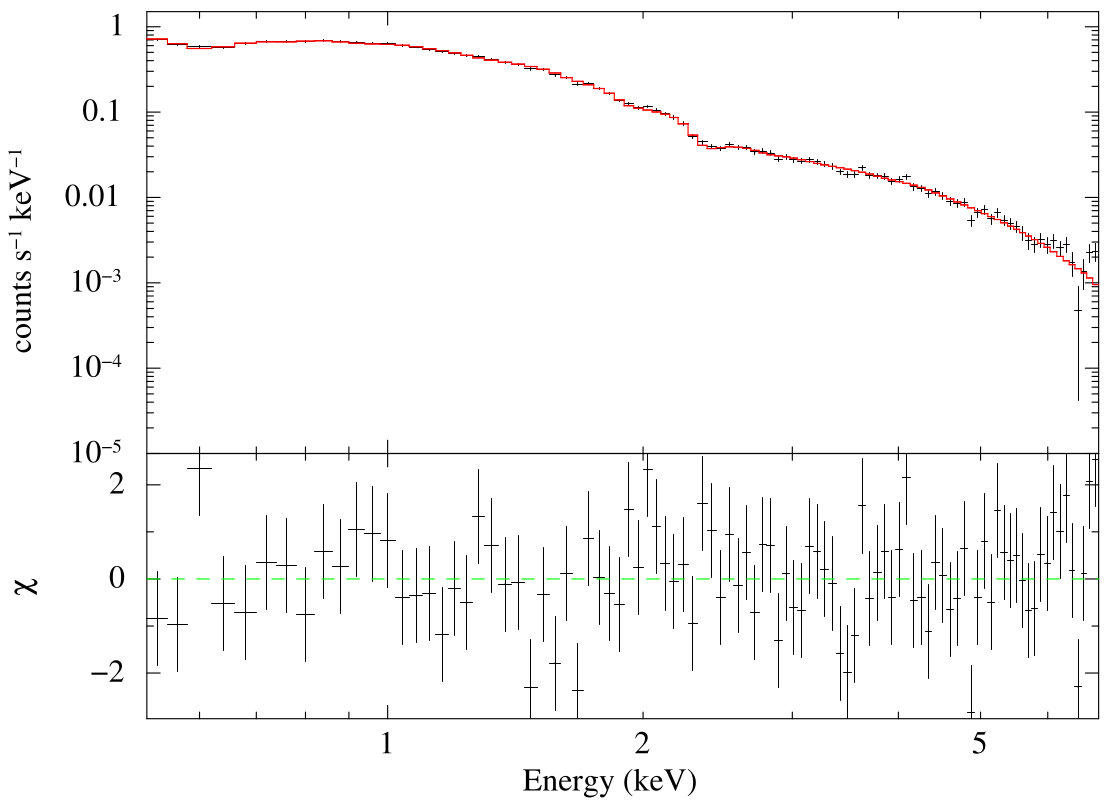}	
\caption{SXT spectrum in the 0.5$-$7 keV energy range fitted with the model $\mathtt {tbabs \times relxillcp}$.} 
\label{figure 6}
\end{figure}

Our analysis of the correlation between the FUV and X-ray flux variations indicates a clear lag, with the FUV emission delayed relative to the X-ray emission. Such behaviour is typically consistent with the disk reprocessing scenario, in which variable X-rays irradiate the accretion disk, producing delayed UV/optical responses. Sources exhibiting this behaviour often display strong reflection signatures in their X-ray spectra. A broadband (0.3$-$50 keV) spectral study of Mrk 1044 by \citet{mallick2018high} revealed a pronounced soft X-ray excess, while \citet{barua2023search} recently confirmed the presence of strong soft X-ray excess in the source, finding that the broadband (0.3$-$25 keV) X-ray spectra can be well described by relativistic reflection scenario. To explore the X-ray spectral characteristics of Mrk 1044 during the epoch of \textit{AstroSat} observations, we performed X-ray spectral fitting of the 0.5$-$7 keV \textit{AstroSat}/SXT spectrum using $\mathtt{ XSPEC}$ (Version 12.15.0d). Prior to spectral fitting, we applied a gain correction, obtaining an offset of 0.030 $\pm$ 0.005 keV, consistent with expected SXT calibration offsets. A systematic uncertainty of 2\% was also added to the SXT spectrum. At first we modeled the spectrum by using a simple absorbed power law model, $\mathtt{tbabs \times powerlaw}$. The fit was statistically poor with a reduced $\chi^{2}$ of 6.7 for 94 degrees of freedom (dof), showing clear residuals in both the soft (0.5$-$1 keV) and hard ($>$4 keV) energy bands. Prominent residuals in the 0.5$-$1 keV energy range indicate a strong soft excess, consistent with earlier studies \citep{mallick2018high,barua2023search}. 
\\[6pt]
Assuming that the observed soft excess arises from relativistic reflection, we employed the physically motivated model $\mathtt{relxill}$ \citep{garcia2014improved} in the form $\mathtt{tbabs \times relxillcp}$. The $\mathtt{relxill}$ model self-consistently combines ionized reflection from the accretion disk ($\mathtt {xillver}$) with relativistic blurring effects ($\mathtt {relline}$) produced by the strong gravitational field near the black hole. Among the various $\mathtt {relxill}$ variants, we adopted $\mathtt {relxillcp}$, which assumes a thermally Comptonized primary continuum ({$\mathtt {nthcomp}$) and allows the disk density to vary in the range of $10^{15}$ $-$ $10^{20}$ cm$^{-3}$.  During the fitting, we allowed the following parameters to vary: emissivity index 1 ($q_{\rm 1}$), inclination angle ($\theta$), black hole spin parameter (a$^{*}$), photon index ($\Gamma$), ionization parameter (log $\xi$), accretion disk density (log $N$), iron abundance ($A_{\rm Fe}$), reflection fraction ($f_{\rm refl}$), and the normalization (\textit{Norm}). The break radius ($R_{\rm br}$) was frozen to a value of 3.0 following previous studies by \cite{mallick2018high} and \cite{barua2023search}, while $kT_{\rm e}$ was fixed to 100 keV. The Galactic hydrogen column density ($N_{\rm H}$) was frozen to a value of 2.9 $\times$ $10^{20}$ cm$^{-2}$, as obtained from the $N_{\rm H}$ calculator available at $\mathtt {HEASARC}$ website\footnote{https://heasarc.gsfc.nasa.gov/cgi-bin/Tools/w3nh/w3nh.pl} which is based on the HI4PI survey \citep{bekhti2016hi4pi}. The remaining model parameters were frozen to their default values during the fit. This model provided an acceptable fit with a reduced $\chi^{2}$ of 1.2 for 87 dof with a null hypothesis probability of $p$ = 0.08. Parameter uncertainties at the 90\% confidence level were estimated using MCMC simulations in $\mathtt {XSPEC}$, employing the Goodman–Weare algorithm \citep{goodman2010ensemble} with 20 walkers and a total chain length of 2$\times$$10^{5}$. The best-fit spectrum with residuals is shown in Fig. \ref{figure 6}, and the corresponding model parameters are listed in Table \ref{Table 1}.
\\[6pt]
From the physical model fit to the observed spectrum, we also estimated the mass accretion rate ($\dot{m}$) defined as the ratio of bolometric luminosity ($L_{\rm bol}$, erg s$^{-1}$) to the Eddington luminosity ($L_{\rm Edd}$, erg s$^{-1}$). To estimate $L_{\rm bol}$ we adopted the following approach. Firstly, using the  $\mathtt {flux}$ command in $\mathtt {XSPEC}$, we measured the observed flux as $F_{\rm 2-10 ~keV}^{\rm obs}$ = $2.40_{-0.11}^{+0.06}$ $\times$ $10^{-11}$ erg s$^{-1}$ cm$^{-2}$ in the 2 $-$ 10 keV energy range, and by applying the convolution model $\mathtt {clumin}$ in the form $\mathtt {tbabs \times clumin \times relxillcp}$, we derived the absorption-corrected intrinsic luminosity as $L_{\rm 2-10 ~keV}^{\rm rest,~abscorr}$ = (1.38$\pm$0.03)} $\times$ $10^{43}$ erg s$^{-1}$ in the 2 $-$ 10 keV energy range. From the absorption-corrected intrinsic luminosity $L_{\rm 2-10 ~keV}^{\rm rest,~abscorr}$, we estimated $L_{\rm bol}$ after applying the bolometric correction from \cite{duras2020universal} as:

\begin{equation}
    L_{\rm bol} = K_{\rm X} \times \rm L_{\rm 2-10 ~keV}
\end{equation}

and $K_{\rm X}$ is 

\begin{equation}
    K_{\rm X} (\rm L_{\rm 2-10 ~keV}) =   a\left[1 + \left(\frac{\rm log_{10}(L_{\rm 2-10~keV}/L_\odot)}{b} \right)^{c}\right] 
\end{equation}
where, $a$ = 15.33 $\pm$ 0.06, $b$ = 11.48 $\pm$ 0.01 and $c$ = 16.20 $\pm$ 0.16.

 \subsection{Comparison of measured and model time lag}
\begin{figure*}
\hbox{
\hspace{1 cm}
\includegraphics[scale=0.45]{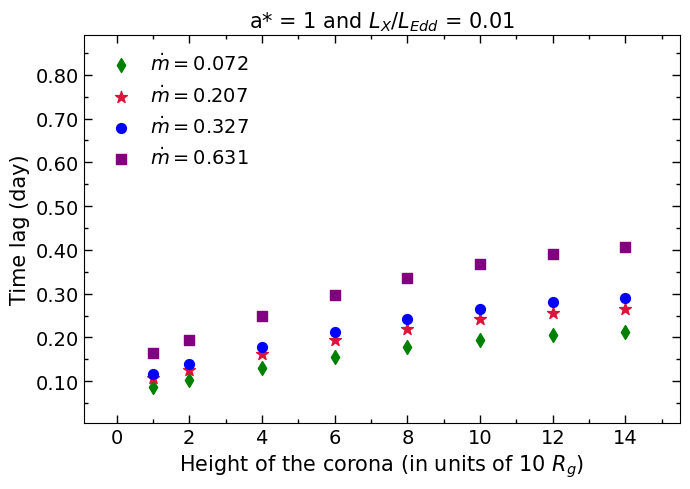}	
\hspace{0.5 cm}
\includegraphics[scale=0.45]{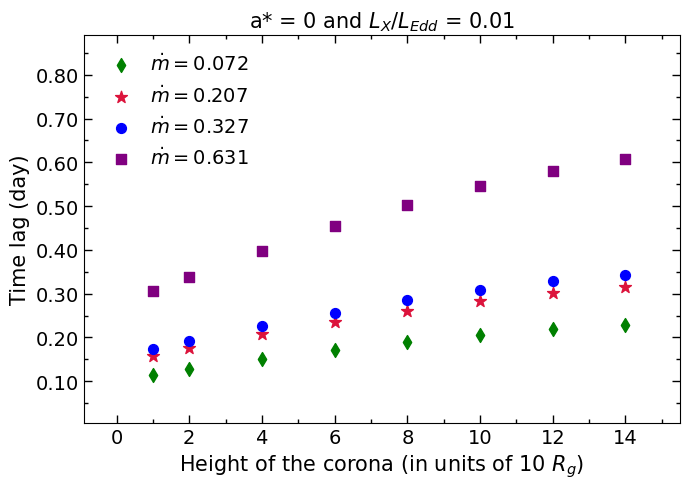}
}
\vspace{0.5 cm}
\hbox{
\hspace{1 cm}
\includegraphics[scale=0.45]{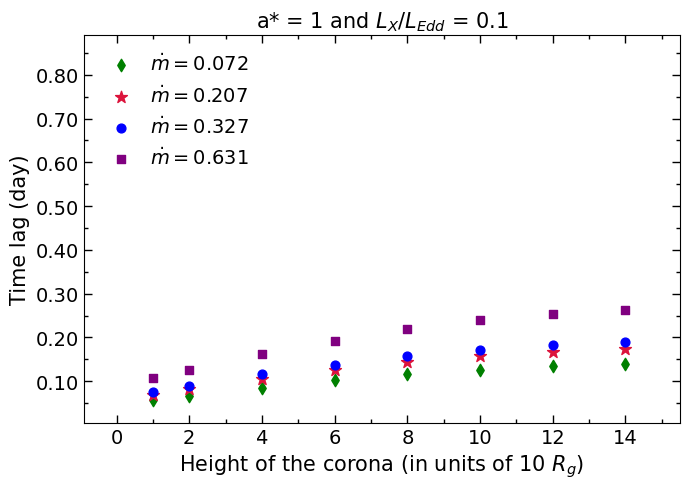}	
\hspace{0.5 cm}
\includegraphics[scale=0.45]{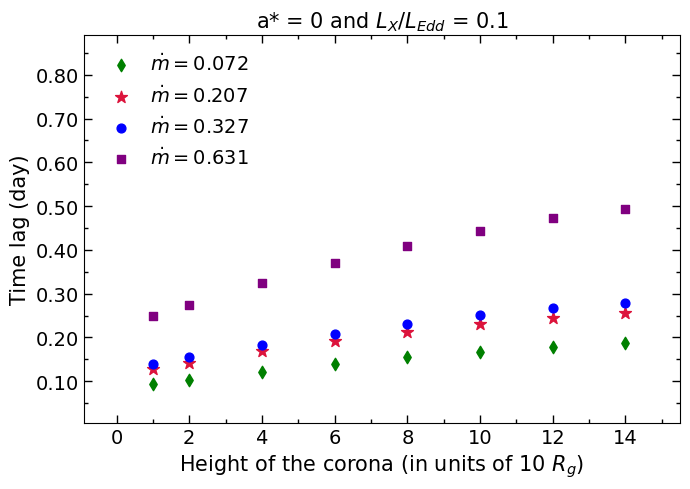}
}
\caption{Model time lag against the height of the corona for different mass accretion rates ($\dot{m}$). In both the top and bottom panels, the left plots are for black hole spin, a$^{*}$ = 1, while the right plots are for black hole spin, a$^{*}$ = 0. Further, the top and bottom panels are for $L_{X}$/$L_{Edd}$ = 0.01 and 0.1 respectively.} 
\label{figure 7}
\end{figure*}

A plausible explanation for the observed increase in time delay with wavelength is thermal reprocessing. This mechanism was recently explored by \citet{kammoun2021uv}, who modelled an X-ray point source illuminating a Novikov–Thorne accretion disk \citep{novikov1973astrophysics,shakura1973black} surrounding a rotating black hole, incorporating both special and general relativistic effects. They derived an analytical expression describing the time lag as a function of wavelength, which depends on the coronal height, black hole mass ($M_{\rm BH}$), mass accretion rate ($\dot{m}$), dimensionless black hole spin parameter (a$^{*}$), and observed X-ray luminosity. The model-predicted time lag (day) can be expressed as \citep{kammoun2021uv}:

\begin{equation}
    \tau_{\rm cen} (\lambda) = A(h_{\rm 10}) M_{\rm 7}^{\rm 0.7} f_{\rm 1}(\dot{m}_{\rm 0.05}) f_{\rm 2}(L_{\rm X,0.01}) \lambda_{\rm 1950}^{B(h_{\rm 10})}
\end{equation}
where, $h_{\rm 10}$ is the height of the corona above the black hole in units of 10 $R_{\rm g}$ ($R_{\rm g}$ = G$M_{\rm BH}$/$c^{2}$, the gravitational radius and G is the universal Gravitational constant and $c$ is the speed of light), $\dot{m}_{\rm 0.05}$ is the mass accretion rate in units of 5\% of the Eddington limit, $M_{\rm 7}$ is the mass of the black hole in units of $10^{7}$ $M_\odot$, $L_{\rm X,0.01}$ is the absorption-corrected intrinsic luminosity in the 2 $-$ 10 keV band in units of 0.01 of the Eddington luminosity and $\lambda_{\rm 1950}$ is the wavelength normalized to 1950 \AA. The full functional forms of A($h_{\rm 10}$), B($h_{\rm 10}$), $f_{\rm 1}$($\dot{m}_{\rm 0.05}$) and $f_{\rm 2}$($L_{\rm X,0.01}$) for both a non-rotating Schwarzschild black hole (a$^{*}$ = 0) and a maximally rotating kerr black hole (a$^{*}$ = 1) are provided in Equations (9 $-$ 16) of \cite{kammoun2021uv}. For Mrk 1044, we used the observed values of $M_{\rm BH}$, the 2 $-$ 10 keV energy range luminosity, and the effective wavelength of the FUV filter ($\lambda$ = 1541 \AA) of UVIT to compute the model-predicted time lag for coronal heights ranging from 1 $R_{\rm g}$ to 150 $R_{\rm g}$. The resulting theoretical lags are shown in Fig. \ref{figure 7} for accretion rates of 0.072, 0.207 and 0.327 reported by \citet{laha2018x}, and for 0.631 $\pm$ 0.014 estimated from our SXT spectral fit (see Fig. \ref{figure 6}). We found that the observed lag between FUV and X-rays is roughly five times larger than that predicted by the analytical model across a wide range of $\dot{m}$, coronal heights, and two values of the absorption-corrected 2 $-$ 10 keV intrinsic luminosity expressed in Eddington luminosity. Such a large discrepancy between the measured and predicted lags cannot be explained solely by uncertainties in the adopted values of $M_{\rm BH}$, $\dot{m}$ or 2 $-$ 10 keV luminosity. The underlying model used here assumes a simplified lamp-post geometry with general relativistic corrections applied. However, the observed soft X-ray excess below $\sim$1 $-$ 2 keV is often attributed to Comptonization in a warm, optically thick ($\tau > 10$), and relatively cool ($kT\sim$0.1 $-$ 1 keV) corona situated on or above the accretion disk \citep{petrucci2018testing}. As a result, part of the radiation that ultimately illuminates the outer, cooler disk regions, and thus drives the UV/optical response, originates from this spatially extended warm corona, rather than directly from a compact hot corona. If the warm corona extends over a larger region than the compact lamp-post source, the effective light-travel path between the variable X-ray emitting region and the UV emitting disk annuli increases. In addition, the warm corona may reprocess and re-emit radiation on thermal or Comptonization timescales, thereby damping fast variability \citep{gardner2017origin,kubota2018physical}. Both of these effects act to increase the observed time lag between the X-ray and UV variations relative to the predictions of the adopted model here. 

\begin{table*}
\centering
\caption{The best-fit spectral parameters obtained using $\mathtt {relxillcp}$ model. Here, $^{f}$ denotes the frozen parameters during the fit.} 
\begin{tabular}{c c c} 
\hline
Model components & Parameters & Value   \\ 
\hline
&& \\
Galactic absorption ($\mathtt {tbabs}$) & $N_{\rm H}$ ($10^{20}$ cm$^{-2}$)      & $2.9^{f}$        \\
Relativistic reflection ($\mathtt {relxillcp}$) & $q_{1}$ & $7.43_{-4.87}^{+1.74}$\\
&  $q_{\rm 2}$ &    $3.0^{f}$\\
& $R_{\rm br}$ ($R_{\rm g}$)  &  $3.0^{f}$ \\
& $\theta$ (deg)  & $36.8_{-7.1}^{+4.1}$   \\
& a$^{*}$ &  $0.92_{-0.10}^{+0.07}$   \\
& $R_{in}$ ($R_{\rm g}$) & $1.0^{f}$  \\
& $R_{out}$ ($R_{\rm g}$) & $400.0^{f}$ \\
& $\Gamma$ &  $2.00_{-0.05}^{+0.04}$  \\
& log $\xi$ & $2.78_{-0.06}^{+0.10}$  \\
& log $N$ (cm$^{-3}$)  & $17.03_{-0.56}^{+0.12}$   \\
& $A_{\rm Fe}$  & $3.37_{-0.61}^{+0.62}$  \\
& $kT_{\rm e}$ (keV)  & $100.0^{f}$   \\
& $f_{\rm refl}$ & $4.94_{-1.76}^{+2.94}$  \\
& Norm ($10^{-5}$) & $5.46_{-1.77}^{+1.50}$  \\
Flux   & log $F_{\rm 2-10 ~keV}^{\rm obs}$ (erg s$^{-1}$ cm$^{-2}$)  & $-10.62_{-0.02}^{+0.01}$ \\
Luminosity & log $L_{\rm 2-10 ~keV}^{\rm rest, ~abscorr}$ (erg s$^{-1}$) & $43.14_{-0.01}^{+0.01}$\\
& $\chi^{2}$/dof &  105.65/87\\
 \hline
\end{tabular}
\label{Table 1}
\end{table*}

\section{Discussion}
\label{sec: discussion}
From the $\sim$5 days simultaneous \textit{AstroSat} observations in X-rays and FUV, we detected a clear correlation between the two bands, with the FUV emission lagging the X-rays by approximately 2 days. The peak correlation coefficient obtained from the cross-correlation analysis is 0.69. In comparison, \citet{edelson2019first}, using a long-term monitoring campaign over $\sim$125 days, reported stronger correlations among optical light curves, while the correlations between X-ray and UV variations were notably weaker, with smaller correlation coefficients. A similar trend, where the correlation strength decreases with increasing wavelength, has been observed in the Seyfert 1 galaxy Mrk 1220 using high-quality \textit{Swift} observations during 2016 $-$ 2017 \citep{peng2025investigating}. Likewise, a weak correlation between UV and X-ray variability has been reported for Fairall 9 \citep{partington2024connecting}. According to \citet{panagiotou2022explaining}, such low correlation coefficients are expected in the case of an X-ray illuminated accretion disk due to the dynamic and variable nature of the X-ray source. Alternatively, \citet{edelson2017swift}, from long-term multiwavelength monitoring of NGC 4151, found that the simple reprocessing model, in which the corona illuminates the disk and the emission is reprocessed into UV/optical bands, fails to explain the observed variability. Instead, they favoured the two-stage reprocessing model proposed by \citet{gardner2017origin}, which includes an additional reprocessing region. In NGC 5548, the X-ray lag deviates from the expected relation of $\tau \propto \lambda^{4/3}$ \citep{cackett2021reverberation}, suggesting a more complex disk–corona coupling. Moreover, some AGN show no measurable correlation between UV and X-ray variability. For instance, the narrow-line Seyfert 1 galaxy IRAS 13224$-$3809 exhibits strong X-ray variability and reflection features indicative of disk illumination; however, no correlation between UV and X-ray variations was detected over a $\sim$40-days monitoring campaign \citep{buisson2018there}.
\\[6pt]
It is also plausible that the observed UV $-$ X-ray correlation is timescale dependent. From long-term monitoring, \citet{breedt2010twelve} reported a strong correlation between optical and X-ray variations on week-to-month timescales, but not on month-to-year timescales. Similar behaviour has been observed in NGC 7469, where the correlation strength varied with the timescale of variability \citep{pahari2020evidence}. Moreover, a given source may exhibit correlated UV $-$ X-ray variability during certain epochs but not during others. Mrk 1044, studied in this work, provides a clear example. Using three long \textit{XMM-Newton} observations and several \textit{Swift} datasets, \citet{barua2023search} found no significant correlation between UV and X-ray flux variations on either short or long timescales. In contrast, our analysis of the $\sim$5 days simultaneous \textit{AstroSat} dataset reveals a clear lag between UV and X-ray variations, indicating a strong short-term coupling between the two emission regions. This is expected, as both Comptonization and disk reprocessing are happening simultaneously and the observed lag may be a mixture of the two. Depending on which process dominates over the other at any given time, the observed lag may be reduced, vanish or show complex dependence with energy. These results suggest that the UV $-$ X-ray relationship in AGN is highly complex and time-dependent, and that a simple disk reprocessing model may be insufficient to explain the diverse variability behaviours observed. Future coordinated multiwavelength campaigns covering a larger AGN sample and a broad range of timescales along with careful modelling will be essential to disentangle the physical mechanisms governing disk-corona interactions.

\section{Conclusions}
\label{sec: summary}
We investigated the UV and X-ray variability properties of Mrk 1044 using simultaneous FUV and X-ray observations obtained with UVIT and SXT onboard \textit{AstroSat}, spanning approximately 5 days. The main findings of this study are summarized below:
\begin{itemize}
    \item [(i)] The source showed clear and significant variability in both the FUV and X-ray bands during the observation period. Using the full duration of the FUV ($\sim$65 ks) and X-ray ($\sim$178 ks) datasets, we measured  $F_{\rm var}$ of 0.036 $\pm$ 0.001 in the FUV band and 0.384 $\pm$ 0.004 in the X-ray band. When considering only the overlapping interval between the FUV and X-ray observations, the corresponding $F_{\rm var}$ values are 0.036 $\pm$ 0.001 and 0.370 $\pm$ 0.004, respectively. Thus, Mrk 1044 is found to be significantly variable in X-rays compared to the FUV band.
    \item [(ii)] From the cross-correlation analysis, we found that the FUV variations lag behind the X-ray variations, consistent with an X-ray reprocessing scenario in this system. However, the correlation likely involves additional complex physical processes, as suggested by the diverse UV $-$ X-ray relationships observed among AGN. This represents the first detection of a measurable time lag in Mrk 1044. Using the ICCF method, we obtained a lag of $\tau$ = 2.25$\pm$0.05 days, while the JAVELIN analysis yielded $\tau$ = $2.35_{-0.01}^{+0.02}$ days.
    \item [(iii)] We compared the observed time lag with predictions from the model of \cite{kammoun2021uv}, which accounts for different coronal heights, accretion rates, and absorption-corrected 2 $-$ 10 keV intrinsic luminosities corresponding to 0.01$L_{\rm Edd}$ and 0.1$L_{\rm Edd}$. The observed lag was found to be significantly larger than the model predictions, which may arise from uncertainties in $M_{\rm BH}$, $\dot{m}$, coronal geometry, and/or the contribution of an extended warm corona. 
\end{itemize}

\section*{Acknowledgments}
The authors thank the anonymous reviewer for the insightful comments and suggestions that significantly improved this manuscript. This study utilized data from the \textit{AstroSat} mission conducted by the Indian Space Research Organization (ISRO), which is archived at the Indian Space Science Data Centre (ISSDC). The work also benefited from the data obtained through the Soft X-ray Telescope (SXT) developed at Tata Institute of Fundamental Research (TIFR), Mumbai, India. The UVIT project represents a collaborative effort involving Indian Institute of Astrophysics (IIA) Bangalore, India, Inter-University Centre for Astronomy and Astrophysics (IUCCA) Pune, India, TIFR Mumbai and various ISRO and CSA centres. The SXT POC at TIFR and UVIT POC at IIA in Bangalore played a crucial role in verifying and releasing the data via the ISSDC data archive. A.K.M. acknowledges the support from the European Research Council (ERC) under the European Union’s Horizon 2020 research and innovation program (grant No. 951549). One of the authors (SBG) thanks IUCAA, Pune, India for the Visiting Associateship.

\section*{Data availability}
The data used in this work are available in the Indian Space Science Data Center at https://www.issdc.gov.in/AS1\_Data\_Download.html

\bibliographystyle{elsarticle-harv} 
\bibliography{ref}
\end{document}